\documentclass[a4paper,11pt]{article}

\usepackage[centertags]{amsmath}
\usepackage{amsfonts}
\usepackage{amssymb}
\usepackage{amsthm}
\usepackage{graphicx}
\usepackage{a4}

\numberwithin{equation}{section}

\newcommand{\Ne}[1]{$\mathcal{N}={#1}$}		

\begin{document}

\thispagestyle{empty}

\begin{flushright}
\small
DFTT 15/2002\\
NORDITA-2002-30 HE\\
hep-th/0205216\\
\normalsize
\end{flushright}

\begin{center}
 

\vspace*{1.5cm}

{\Large {\bf On the $\mathcal{N}=1\,$ $\beta$-function from the conifold \footnote{Work
partially supported by the European Commission Marie Curie Training Site
Fellowship HPMT-CT-2000-00010 and RTN programme HPRN-CT-2000-00131.}}}

\vspace{2cm}

{\bf Emiliano Imeroni}\\

\vskip 1cm
\emph{Dipartimento di Fisica Teorica, Universit\`a di Torino\\
and I.N.F.N., Sezione di Torino, Via P. Giuria 1, I-10125 Torino, Italy}
\vskip 0.3cm
\emph{and}
\vskip 0.3cm
\emph{NORDITA,
Blegdamsvej 17, DK-2100 Copenhagen \O, Denmark} \\
\vskip 0.4cm
email: \verb|imeroni@to.infn.it|\\

\end{center}

\vspace{3cm}

\begin{abstract}
We obtain the correct all-loop $\beta$-function of pure \Ne{1} super Yang--Mills theory from the supergravity solution of the warped deformed conifold, including also some nonperturbative corrections. The crucial ingredient is the gauge-gravity relation that can be inferred by taking into account the phenomenon of gaugino condensation.
\end{abstract}

\newpage

\section{Introduction}

In the search for non conformal extensions of the gauge-gravity correspondence, the case of \Ne{1} super Yang--Mills theory in four spacetime dimensions has turned out to be an especially interesting and fruitful arena. In this case it has proved possible to construct nonsingular supergravity solutions which are in principle able to give information on the dual quantum gauge theory at all energy scales.

Two supergravity backgrounds dual to \Ne{1} gauge theories are:
\begin{itemize}
\item the warped deformed conifold by Klebanov and Strassler (KS) \cite{Klebanov:2000hb}, describing a deformation of the geometry produced by regular and fractional D3-branes at a conifold singularity;
\item the Maldacena--Nu\~nez (MN) solution \cite{Maldacena:2000yy}, describing D5-branes wrap\-ped on a two-cycle inside a Calabi--Yau threefold.
\end{itemize}

An important result one can obtain from these backgrounds is the running of the coupling constant of the dual gauge theory. The one-loop $\beta$-function of the $SU(N+M)\times SU(N)$ gauge theory which is dual to the conifold solution was obtained in \cite{Herzog:2001xk}, by identifying the energy scale at which the gauge theory is defined with a radial coordinate in the classical solution. In order to go beyond the one-loop approximation, different gauge-gravity relations were introduced, but none of them allowed to go further and reproduce the correct ratio between the one- and two-loop coefficients of the $\beta$-function.

The reason of this can be traced to the fact that in \cite{Herzog:2001xk} only the UV asymptotic behavior of the solution was considered. However, although the perturbative running of the coupling constant is of course a UV phenomenon, the exact gauge theory scale $\Lambda$ is generated in the IR by dimensional transmutation, after the theory has flown through a duality cascade to a theory which is supposed to lie in the same universality class of pure $SU(M)$ SYM. Thus, the identification of the correct relation between gauge theory scale and supergravity quantities has necessarily to take into account the IR details of the deformed KS solution. 

In this brief note we show, following closely related results obtained in \cite{DiVecchia:2002??} by Di Vecchia, Lerda and Merlatti (DLM) in the context of the MN solution, how it is possible to obtain the $\beta$-function of pure \Ne{1} SYM theory at \emph{all-loops} from the KS solution, by implementing the gauge-gravity relation that one obtains when taking into account the phenomenon of gaugino condensation in the gauge theory. In addition, by a slight refinement of the analysis we get exponential corrections to the running, which have the form of fractional instanton contributions. The general result can also be seen as a quantitative confirmation of the fact that the gauge theory which lies at the end of the duality cascade is effectively pure $SU(M)$ SYM.

This note is organized as follows. In section \ref{review} we briefly review the elements of the KS solution that are needed for our derivation. Next, in section \ref{gaugegra} we discuss the gauge-gravity relation that is obtained by identifying which supergravity quantity is dual to the gaugino condensate in the gauge theory. Finally, in section \ref{betafunc} we compute the $\beta$-function of the theory.

\section{Brief review of the KS solution}\label{review}

Consider the geometry generated by $N$ regular D3-branes and $M$ fractional D3-branes (namely, D5-branes wrapped on a vanishing two-sphere) at the apex of a conifold. The dual gauge theory is a non conformal \Ne{1} SYM theory in four spacetime dimensions with gauge group $SU(N+M)\times SU(N)\,$, coupled to two chiral superfields transforming in the $(\mathbf{N+M},\mathbf{\bar{N}})$ representation and two transforming in its conjugate, with an appropriate superpotential. If $N$ is a multiple of $M$, then this theory flows to $SU(M)$ in the IR, via a chain of Seiberg dualities known as ``duality cascade'' which reduces the rank of the gauge groups by $M$ units at each cascade jump. As a result, $M$ units of R-R 3-form flux are turned on, blowing up an $S^3$ in the geometry, so that the conifold is replaced by the deformed conifold \cite{Klebanov:2000hb}. The deformation of the conifold can be seen as the translation in supergravity of the chiral symmetry breaking of the $SU(M)$ gauge theory.

The metric takes the form:
\begin{equation}\label{metric}
	ds^2_{10} =   h^{-1/2}(\tau)   dx_n dx_n
	 	+  h^{1/2}(\tau) ds_6^2 \,,
\end{equation}
where the function $h(\tau)$ is given by the integral expression:
\begin{equation}
	h(\tau) = 
		(g_s M\alpha')^2 2^{2/3} \varepsilon^{-8/3} 
		\int_\tau^\infty d x \frac{x\coth x-1}{\sinh^2 x} (\sinh (2x) - 2x)^{1/3}\,,
\end{equation}
where $\varepsilon$ is the deformation parameter of the conifold. In the basis given by some appropriate one-forms\footnote{We will not need the explicit expression of the one-forms $g^i$ here. It was found in \cite{Minasian:1999tt}. For an easy reference see \cite{Herzog:2001xk}.}, the metric on the six-dimensional deformed conifold is given by:
\begin{multline}
	ds_6^2 = \frac{1}{2}\varepsilon^{4/3} K(\tau)
		\Bigg[ \frac{1}{3 K^3(\tau)} (d\tau^2 + (g^5)^2)
		+ \cosh^2 \left(\frac{\tau}{2}\right) [(g^3)^2 + (g^4)^2] \\
		+ \sinh^2 \left(\frac{\tau}{2}\right)  [(g^1)^2 + (g^2)^2] \Bigg]\,,
\end{multline}
where
\begin{equation}
	K(\tau)=\frac{ (\sinh (2\tau) - 2\tau)^{1/3}}{2^{1/3} \sinh \tau}\,.
\end{equation}
The other bosonic fields present in the solution are a constant dilaton, the NS-NS two-form $B_2\,$ and the R-R potentials $C_2$ and $C_4\,$:
\begin{align}
	B_2 &= \frac{g_s M \alpha'}{2} [f g^1\wedge g^2
		+  k g^3\wedge g^4 ]\,,\\
	F_3 & = dC_2 = \frac{M\alpha'}{2} \left \{ (1- F) g^5\wedge g^3\wedge g^4 
		+ F g^5\wedge g^1\wedge g^2  \right. \nonumber \\
		&\qquad \qquad\left. + F' d\tau\wedge
		(g^1\wedge g^3 + g^2\wedge g^4) \right \}\,,\label{Cdef}\\
	\tilde F_5 &= dC_4 + B_2 \wedge F_3 = \mathcal{F}_5 + \star \mathcal{F}_5\,,
\end{align}
where:
\begin{equation}
	\mathcal{F}_5 = B_2\wedge F_3 = \frac{g_s M^2 {\alpha'}^2}{4} [f(1-F) + k F]
		g^1\wedge g^2\wedge g^3\wedge g^4\wedge g^5\,,
\end{equation}
and:
\begin{align}
F(\tau) &= \frac{\sinh \tau -\tau}{2\sinh\tau}\,, \\
f(\tau) &= \frac{\tau\coth\tau - 1}{2\sinh\tau}(\cosh\tau-1) \,, \\
k(\tau) &= \frac{\tau\coth\tau - 1}{ 2\sinh\tau}(\cosh\tau+1)\,.
\end{align}

\section{Gauge-gravity relation}\label{gaugegra}

One of the problems in extending the gauge/gravity correspondence to non conformal cases resides in the fact that the functional form of the relation between the scale of the gauge theory and the quantities which appear in the supergravity solution is not uniquely determined \cite{Peet:1998wn}.

DLM have shown that the correct gauge-gravity relation in the case of pure \Ne{1} SYM theory can be inferred by taking into account the phenomenon of chiral symmetry breaking \cite{DiVecchia:2002??}. To be specific, one has to find which quantity in the classical solution is dual to the \emph{gaugino condensate}, which is a protected quantity in the gauge theory.

Let us then consider how we can see the breaking of the $U(1)$ R-symmetry of the gauge theory from the point of view of the classical solution. The chiral anomaly can be seen from the variation of the $\theta$-angle of the gauge theory under R-symmetry transformations. The $\theta$-angle is related through the gauge/gravity correspondence to the integral of the supergravity field $C_2$ on the (vanishing) two-sphere $S^2$ of the conifold geometry. The $U(1)$ R-symmetry is simply translated in supergravity into the transformation of an appropriate angle, under which $C_2$ is not invariant, thus breaking the symmetry in the appropriate way down to $\mathbb{Z}_{2M}$ \cite{Klebanov:2002gr,Bertolini:2002xu}. This breaking is a UV phenomenon, and can be seen from the asymptotic form of the classical solution.

Instead, the additional breaking of the symmetry down to $\mathbb{Z}_2$ is a nonperturbative effect due to gaugino condensation in the gauge theory, and is mapped to the deformation of the conifold. An essential point is that this effect takes place after the theory has flown through the duality cascade to a gauge theory which is effectively pure $SU(M)$ SYM.

Anyway, from the considerations above we can derive that, in order to identify what quantity in the classical solution is dual to the gaugino condensate, we can look at the variation of the expression of $C_2$ in the complete solution with respect to its asymptotic form in the UV, where the $\mathbb{Z}_{2M}$ invariance is unbroken.

Let us first see how things work in the case of the MN solution. The dependence on the radial coordinate $\rho$ of the difference of $C_2$ from its asymptotic form at large $\rho$ is given by:
\begin{equation}\label{C2MN}
	\delta C_2^{\text{(MN)}}\sim a(\rho) 
		= \frac{2\rho}{\sinh 2\rho}\,.
\end{equation}
Therefore, the function $a(\rho)$ is responsible of the additional R-symmetry breaking down to $\mathbb{Z}_2\,$, and thus plays the role of the gaugino condensate. This result was also obtained in \cite{Apreda:2001qb} with different techniques. In the gauge theory, the gaugino condensate is a protected operator which does not acquire an anomalous dimension, and is thus fixed by dimensional analysis to be $\langle \lambda \lambda \rangle = c\Lambda^3\,,$ where $\Lambda$ is the exact dynamically generated scale of the theory and $c$ a computable constant. The dimensionless function $a(\rho)$ can then be put in correspondence with the gaugino condensate measured in units of the arbitrary scale $\mu$ at which the theory is defined, and then the gauge-gravity relation between supergravity quantities and the gauge theory scale can be established in the following way:
\begin{equation}\label{ggMN}
	a(\rho)\sim\frac{\Lambda^3}{\mu^3}\,.
\end{equation}
Using the relation~\eqref{ggMN}, it has been shown in \cite{DiVecchia:2002??} that one is able to get the complete $\beta$-function of pure \Ne{1} SYM theory from the MN solution.

Let us consider now the case of the conifold. From~\eqref{Cdef} and from the fact that $F(\tau)\sim\tfrac{1}{2}$ for large $\tau$, we can see (as first computed in \cite{Loewy:2001pq}) that the $\tau$-dependence of the variation of $C_2$ analogous to~\eqref{C2MN} is given by:
\begin{equation}\label{C2}
	\delta C_2\sim \frac{1}{2}-F(\tau) = \frac{\tau}{2\sinh\tau}\,,
\end{equation}
so that, following the same reasoning which led to~\eqref{ggMN}, we can impose the following gauge-gravity relation between $\tau$ and the scales of the gauge theory:
\begin{equation}\label{gg}
	\frac{\tau}{2\sinh\tau} \sim \frac{\Lambda^3}{\mu^3}\,.
\end{equation}

\section{The $\beta$-function}\label{betafunc}

Once established the gauge-gravity relation~\eqref{gg} we can compute the running of the coupling constant in the gauge theory. Let us stress again that the relation~\eqref{gg} has been found by using the fact that the theory exhibits chiral symmetry breaking, which is essentially related to the deformation of the conifold. Therefore, we expect to be working at the point where the duality cascade has stopped and the gauge theory at hand is effectively pure \Ne{1} $SU(M)$ SYM.

In principle, it is known that it is problematic to define an exact gauge-gravity duality away from the far IR of this gauge theory, because of a mismatch between the regime in which the supergravity approximation is reliable and the one in which the massive KK modes coming from the blown-up three-sphere decouple and the cascade jumps are well separated \cite{Klebanov:2000hb}. An analogous situation holds for the MN solution \cite{Maldacena:2000yy}. However, establishing an exact duality is beyond the purpose of this paper, while our interest resides in the use of the supergravity solution for extracting information on the gauge theory. With our reduced goal, we can consider small values of $g_sM$ in order to have pure \Ne{1} SYM theory at all regimes, the small and large $\tau$ regions of the solution describing respectively the IR and UV regimes of the gauge theory. We will show that it is possible to extract relevant information on the whole UV regime of the pure SYM theory for any number of colors $M$, and one is able to reproduce the precise running of the coupling constant at all loops, even including some nonperturbative corrections, exactly as in the case of the MN solution \cite{DiVecchia:2002??}.

We can also consider this result as a confirmation of the fact that the asymptotically free and confining theory which is dual to the warped deformed conifold at the end of the RG cascade lies indeed, in a precise quantitative sense, in the universality class of pure \Ne{1} $SU(M)$ SYM.

In the regime we are considering, the effective degrees of freedom of the supergravity configuration are the $M$ D5-branes wrapped on the collapsed two-cycle $S^2$ of the conifold geometry. The coupling constant of the $SU(M)$ gauge theory can be extracted from the quantities of the classical solution through the following relation, which has been presented in different forms in a variety of contexts and we use here in the form introduced in \cite{DiVecchia:2001uc}:
\begin{equation}\label{genrun}
	\frac{1}{g^2(\mu)}=\frac{V_{\text{ST}}\left(S^2\right)}{g^2_{\text{D}3}}
\end{equation}
where $g^2_{\text{D}3}=4\pi g_s$ and the ``stringy volume'' of the 2-cycle $S^2$ on which the branes are wrapped is given in general by:
\begin{equation}\label{stringy}
	V_{\text{ST}}\left(S^2\right)=
		\frac{1}{(2\pi\sqrt{\alpha'})^2}
		\int d^2\zeta \sqrt{\det\left(\mathcal{G}_{AB}+
		B_{AB}\right)}\,.
\end{equation}
where $\mathcal{G}_{AB}$ and $B_{AB}$ are respectively the (unwarped) metric and NS-NS two-form on the cycle. In the present case, one has $\mathcal{G}=0\,$, since the metric seen by the wrapped D5-branes has no support on the vanishing 2-cycle of the geometry.\footnote{Notice that one can get the same result~\eqref{genrun} by expanding the whole world-volume action of a fractional D3-brane in the background generated by the KS solution and studying the coefficient of the kinetic term of the gauge field, in a similar way as it was done in \cite{DiVecchia:2002??}. From the computation one can also see that the warp factors are canceled by the dilaton contribution, and this is the reason why the metric appearing in the ``shortcut'' formula~\eqref{stringy} is unwarped. For a derivation of the fractional D3-brane action from the one of a wrapped D5-brane see for instance appendix B of \cite{DiVecchia:2001uc}.}
The large $\tau$ UV behavior of the $B$-field is given by:
\begin{equation}\label{bfield}
	B_2=\frac{g_sM\alpha'}{2}\ \tau\ \omega_2\,,
\end{equation}
where $\omega_2=\tfrac{1}{2}(g^1\wedge g^2+g^3 \wedge g^4)\,$. Recalling that $\int_{S^2}\omega_2=4\pi$ \cite{Herzog:2001xk}, from~\eqref{genrun}-\eqref{bfield} we get that the gauge coupling is given in terms of the coordinate $\tau$ by:
\begin{equation}\label{gtau}
	\frac{1}{g^2(\mu)}=\frac{M\tau}{8\pi^2}\,.
\end{equation}
From~\eqref{gtau} we can then read the expression of the $\beta$-function as a function of the coordinate $\tau$:
\begin{equation}\label{beta1}
	\beta(g)=\frac{dg}{d\ln\tfrac{\mu}{\Lambda}}
		=-\frac{Mg^3}{16\pi^2}\frac{d\tau}{d\ln\tfrac{\mu}{\Lambda}}
\end{equation}

The final step requires the gauge-gravity relation we have established in the previous section. We first consider the large $\tau$ UV region, where the relation~\eqref{gg} becomes:
\begin{equation}
	\tau e^{-\tau}\sim\frac{\Lambda^3}{\mu^3}\,,
\end{equation}
from which we get:\footnote{Note that it is clear that the precise numerical coefficient entering the relation~\eqref{gg} (or~\eqref{C2}) is not needed for the computation at hand.}
\begin{equation}\label{dtau}
	\frac{d\tau}{d\ln\tfrac{\mu}{\Lambda}}=3\left(1-\frac{1}{\tau}\right)^{-1}
		= 3\left(1-\frac{Mg^2}{8\pi^2}\right)^{-1}\,,
\end{equation}
where we have used~\eqref{gtau}. From~\eqref{dtau} we can finally extract the correct all-loop NSVZ perturbative $\beta$-function of \Ne{1} SYM with gauge group $SU(M)$ \cite{Novikov:uc}:
\begin{equation}\label{beta}
	\beta(g)
		=-\frac{3Mg^3}{16\pi^2}\left(1-\frac{Mg^2}{8\pi^2}\right)^{-1}\,.
\end{equation}
It is quite remarkable that a classical supergravity computation is able to reproduce a quantum property of the dual gauge theory at all orders in perturbation theory, including all numerical factors.

There is additional information to be extracted, related to nonperturbative effects in the gauge theory. If, instead of keeping only the large $\tau$ asymptotics as done before, we use the complete gauge-gravity relation~\eqref{gg}, \eqref{dtau} gets modified as follows:
\begin{equation}
	\frac{d\tau}{d\ln\tfrac{\mu}{\Lambda}}=3\left(1-\frac{1}{\tau}
		+\frac{2e^{-2\tau}}{1-e^{-2\tau}}\right)^{-1}
		= 3\left(1-\frac{Mg^2}{8\pi^2}
		+\frac{2e^{-16\pi^2/Mg^2}}{1-e^{-16\pi^2/Mg^2}}\right)^{-1}\,,
\end{equation}
and the $\beta$-function reads accordingly:
\begin{equation}\label{betanp}
	\beta(g)
		=-\frac{3Mg^3}{16\pi^2}
		\left(1-\frac{Mg^2}{8\pi^2}
		+\frac{2e^{-16\pi^2/Mg^2}}{1-e^{-16\pi^2/Mg^2}}\right)^{-1}\,.
\end{equation}
The exponential corrections in~\eqref{betanp} seems to be due to nonperturbative effects in the gauge theory, having the form of contributions of instantons with fractional charge $2/M\,$. Exactly the same result has been found from the MN solution in \cite{DiVecchia:2002??}, thus strengthening its validity for pure \Ne{1} SYM theory. Although it is generally believed that the NSVZ $\beta$-function does not receive nonperturbative corrections, the arguments leading to such a conclusion do not take into account the possible contributions of the monopole-type configurations discussed in \cite{Davies:1999uw}, which are supposed to be responsible of gaugino condensation. These configurations have indeed fractional instanton charge. It would be of great interest to study these issues from the point of view of the pure gauge theory and to compare the behavior and the precise numerical coefficients with the predictions given by the supergravity analysis.

\subsection*{Acknowledgments}

I am grateful to P.~Di Vecchia, A.~Lerda and P.~Merlatti for having shared and discussed with me their ideas and results and for comments on this work. I would also like to thank M.~Bertolini for discussions and encouragement.


\end{document}